# Intermolecular charge transfer enhances the performance of molecular rectifiers

Ryan P. Sullivan[1], John T. Morningstar[2], Eduardo Castellanos[1], Robert W. Bradford III[1], Yvonne J. Hofstetter,[3,4] Yana Vaynzof,[3,4] Mark E. Welker[2], Oana D. Jurchescu[1*]

[1]Deparment of Physics, Center for Functional Materials, Wake Forest University, Winston-Salem, NC, 27109, USA
[2]Deparment of Chemistry, Center for Functional Materials, Wake Forest University, Winston-Salem, NC, 27109, USA
[3] Integrated Centre for Applied Physics and Photonic Materials, Technische Universität Dresden, Nöthnitzer Str. 61, 01187 Dresden, Germany
[4] Center for Advancing Electronics Dresden (cfaed), Technische Universität Dresden, Helmholtzstraße 18, 01089 Dresden, Germany

Molecular-scale diodes made from self-assembled monolayers (SAMs) could complement silicon-based technologies with smaller, cheaper, and more versatile devices. However, advancement of this emerging technology is limited by insufficient electronic performance exhibited by the molecular current rectifiers. We overcome this barrier by exploiting the charge-transfer state that results from co-assembling SAMs of molecules with strong electron donor and acceptor termini. We obtain a substantial enhancement in current rectification, which correlates with the degree of charge transfer, as confirmed by several complementary techniques. These findings provide a previously unexplored method for manipulating the properties of molecular electronic devices by exploiting donor/acceptor interactions. They also serve as a model test platform for the study of doping mechanisms in organic systems. Our devices have the potential for fast widespread adoption due to their low-cost processing and self-assembly onto silicon substrates, which could allow seamless integration with current technologies.

**Introduction**

The growing demands for smaller and faster electronics have sparked the need for new componentry technologies that could continue the astonishing pace of the technological advancement witnessed by the semiconductor industry in the past decades. New devices must be compatible with the existing infrastructure and provide functionalities at length scales currently impossible; thus, they should exhibit a reduced footprint. The emerging field of molecular electronics is seeking to address this challenge by using single molecules or self-assembled monolayers (SAMs) of organic molecules as the fundamental building blocks of electric functionality. The exceptional chemical versatility of these molecules allows for tailoring of a vast range of properties, providing opportunities for incorporation in various optoelectronic applications such as diodes, transistors, sensors, or memory switches (*1–8*). One of the simplest devices, the molecular rectifier, has been at the forefront of this research since Aviram and Ratner proposed the first molecular-scale equivalent of a silicon p-n junction diode in 1974 (*9*). Similar to the case of commercial devices, the molecular rectifier induces an asymmetry between the current measured under positive and negative bias regimes, with the figure of merit that quantifies this behavior being the rectification ratio $R = |J(+V)|/|J(-V)|$, where $J$ is current density at the applied voltage $V$. After almost 50 years of research in which various charge transport mechanisms, as well as molecular and device structures have been proposed, unfortunately, the performance of molecular rectifiers remains much inferior to that of silicon diodes, with the highest reported $R$ values around $10^3$, with the exception of one report demonstrating $10^5$ (*1, 3, 10–12*). For reference, $R = 10^5$ to $10^8$ is typical in state-of-the-art commercial diodes. Another drawback is that the majority of these devices are formed on metallic substrates such as gold, silver, or platinum due to the intrinsic challenges associated with the bond formation energies of self-assembly on silicon. Hence, not only do they require complex multistep expensive processing and postprocessing, but their integration with current silicon-dominated technologies would complicate the manufacturing even further and add to the final cost.

Here, we introduce a powerful method to enhance the performance of molecular rectifiers that relies on molecular-level dilution. We demonstrate high-performance molecular diodes by co-assembling SAMs of molecules with electron donor (D) and electron acceptor (A) termini, while also circumventing the high fabrication costs by self-assembling directly onto silicon. Several different molecules, where one can rectify current and the other cannot, were obtained from a simple and high-yield synthetic procedure and co-assembled onto silicon wafers to form binary mixed SAMs; the resulting devices exhibited rectification ratios approaching $10^4$. Our methodology also exploits the fact that silicon has a very low natural roughness and thus requires no additional processing, while empowering seamless integration with other more mature silicon technologies that are currently included in consumer electronics. We show that the charge transfer (CT) between the additive and rectifying molecules, evidenced by ultraviolet photoemission spectroscopy (UPS) measurements, is responsible for the observed increase in the rectification ratio in the mixed SAM devices, despite the reduction caused in the film degree of order. We then demonstrate that these devices can be used in circuits to effectively rectify an AC signal with low ripple voltages. The CT state generated at the donor/acceptor interfaces is known to profoundly impact the properties of devices such as light-emitting diodes, solar cells, or thermoelectrics through various doping schemes and can often lead to novel behavior inaccessible with the parent compounds alone (*13–15*). Doping has been transformational for the conventional semiconductor industry and now serves as a ubiquitous tool for controlling many optoelectronic properties in modern electronics. Its touted potential also prompted attention from the organic semiconductor research (*14*, *16–18*); however, to the best of our knowledge, this is the first time it has been adopted in molecular electronics. Our results uncover new strategies for manipulating the properties of molecular electronic devices by exploiting the hybrid electronic states at the interface between charge-donor and charge-acceptor SAMs. The SAM co-assembly studied here provides an excellent model system that could be further exploited as a simple test platform for the study of basic mechanisms of doping in organic systems. The resulting knowledge could guide the design of more complex systems that rely on the CT between donor and acceptor units, thus expanding the impact to other related fields.

**Results**

We fabricated molecular tunnel junctions based on SAMs placed on highly doped Si substrates, which also served as bottom electrode, and a liquid metal EGaIn (eutectic alloy of gallium and indium) top electrode. The molecules in each SAM consist of two asymmetric anchoring moieties connected with an electrically decoupling sigma-bonded bridge: The head of each molecule is a triethoxysilane designed to covalently bond to the $SiO_2$ (silicon oxide) layer natively formed at the surface of the silicon substrates via covalent Si—O bonds, while the tail consists of a substituted benzene ring that physisorbs to the $Ga_2O_3$ (gallium oxide) that instantaneously forms at the surface of the EGaIn electrode in air. The noninvasive soft top electrode ensures a near 100% yield in the device characterization. Several different substituents have been tested, as described in Materials and Methods; the chemical formula of (*E*)-1-(4-carbomethoxy-phenyl)-*N*-(3-(triethoxysilyl)propyl)methanimine (CMPTM), where the substituent is (—$COOCH_3$), is included in Fig. 1A (left). The chemical formulas corresponding to the other rectifying molecules are shown in fig. S1. These compounds were both tested as single-component SAMs, and diluted systematically with (3-aminopropyl)triethoxysilane (APTES; Fig 1A, right), which was co-assembled using the co-adsorption method in concentrations ranging from 0 to 100%. Figure 1B displays typical current-voltage curves for molecular diodes based on pure CMPTM (red), pure additive APTES (blue), and a mixture of the two (gray), while in Fig. 1C, we show the evolution of the rectification ratio *R* with the APTES content; the results correspond to a bias of ±2 V and are normalized with respect to the average value in the absence of APTES; the dependence of *R* on the applied voltage is shown in fig. S2. The devices fabricated on pure CMPTM SAMs exhibited average *R* values of 1000 ± 300, which is on par with the performance of the best molecular rectifiers fabricated on silicon substrates (Fig. 1D) (*11*, *19*). The performance of mixed SAM films is highly sensitive to the concentration of the additive: As the APTES concentration increases, the current rectification becomes more efficient, with a peak being observed around 30 to 50% APTES, as illustrated in Fig. 1C. At 30% APTES, the average *R* is

4300 ± 1000 (histogram included in Fig. 1E), a more than four times increase compared to the devices based on undiluted CMPTM, with the maximum value reaching 8500. To the best of our knowledge, this represents the most efficient current rectification recorded in molecular rectifiers fabricated on silicon substrates. Such a notable increase does not result from the current rectification of the additive molecule since APTES is very inefficient, yielding an average value of $R = 270 \pm 180$ (Fig. 1F). Rather, it appears that the co-assembly of the two SAMs is necessary to enhance rectification strength of the molecular diode. Similar results have been obtained on other additive/molecular rectifier combinations, with various levels of rectification enhancement; data are shown in figs. S3 and S4. These films of mixed SAMs yield current rectifications exceeding that of the parent compounds, suggesting that our new approach to controlling the properties of molecular diodes is versatile and powerful. In addition, it appears that a minor improvement in $R$ is obtained upon bias stressing the device based on SAM co-deposition (fig. S5), an effect that will be investigated in more detail in a different study. Once the saturation value of $R$ has been reached, the samples appear to be stable upon further stressing.

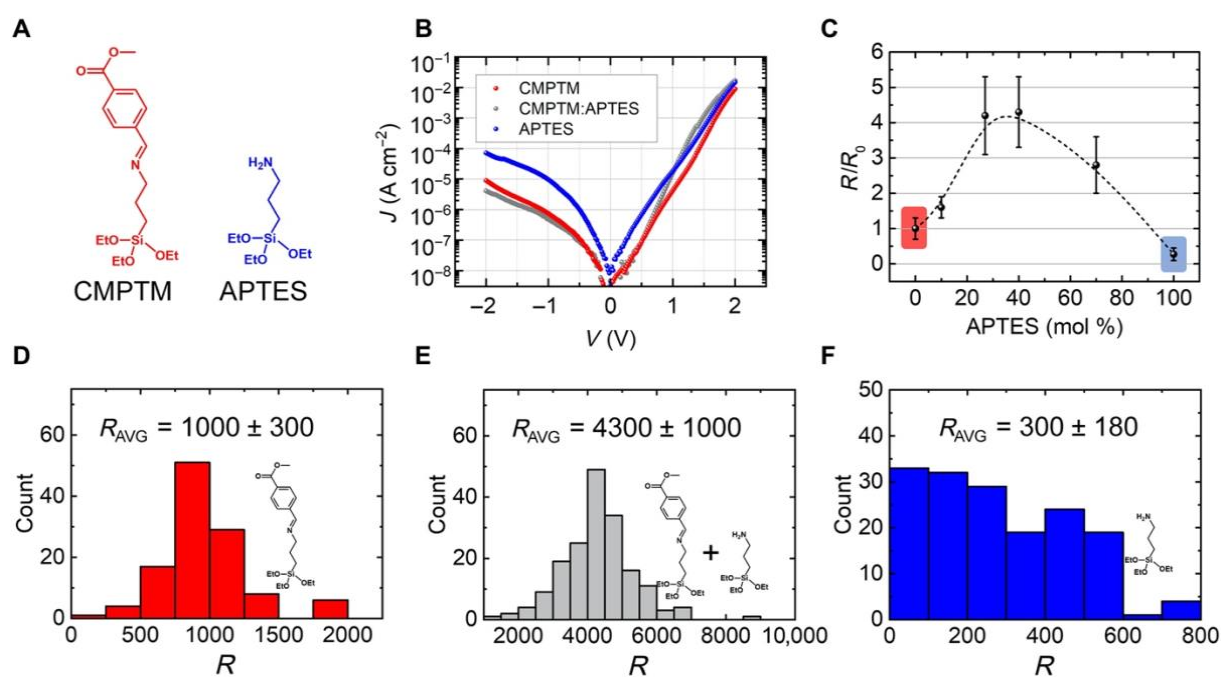

**Fig. 1. Current rectification in CMPTM:APTES monolayers.** (**A**) Molecular structures of the rectifying molecule CMPTM (red) and the additive molecule APTES (blue). (**B**) Current density versus applied voltage in molecular tunnel junctions consisting of SAMs of CMPTM (red), CMPTM co-assembled with APTES at 70:30 (gray), and APTES (blue). (**C**) Rectification ratio $R$ versus concentration of APTES in mixed rectifying SAMs of CMPTM (the value is normalized with respect to the average rectification obtained in pure CMPTM devices, $R_0$). (**D**) Histogram of rectification ratios measured in pure CMPTM devices. (**E**) Histogram of rectification ratios measured in CMPTM co-assembled with APTES at 70:30. (**F**) Histogram of rectification ratios measured in pure APTES devices. The average rectification value is included in the inset for (D) to (F).

To understand the mechanism responsible for this enhancement in device performance and be able to expand it to other systems, we designed and executed a series of additional experiments.

First, we evaluated if the presence of the additive SAM resulted in an increase in the degree of order in the rectifying monolayer, which, in turn, would yield to higher $R$ (20). Two well-known signatures of a highly ordered, densely packed SAM are the uniformity of the current-voltage curves and a large breakdown voltage (1, 2). Figure 2 presents these parameters measured over multiple spots on the same film of rectifier SAM (left) and mixed SAM (right) samples. A schematic of the composition is illustrated in Fig. 2 (A and B). The current density ($J$) versus voltage ($V$) data are sorted into two-dimensional bins for pure SAMs of CMPTM (Fig. 2C) and mixed SAMs of CMPTM and APTES (Fig. 2D), along with the corresponding breakdown voltages (Fig. 2, E and F).

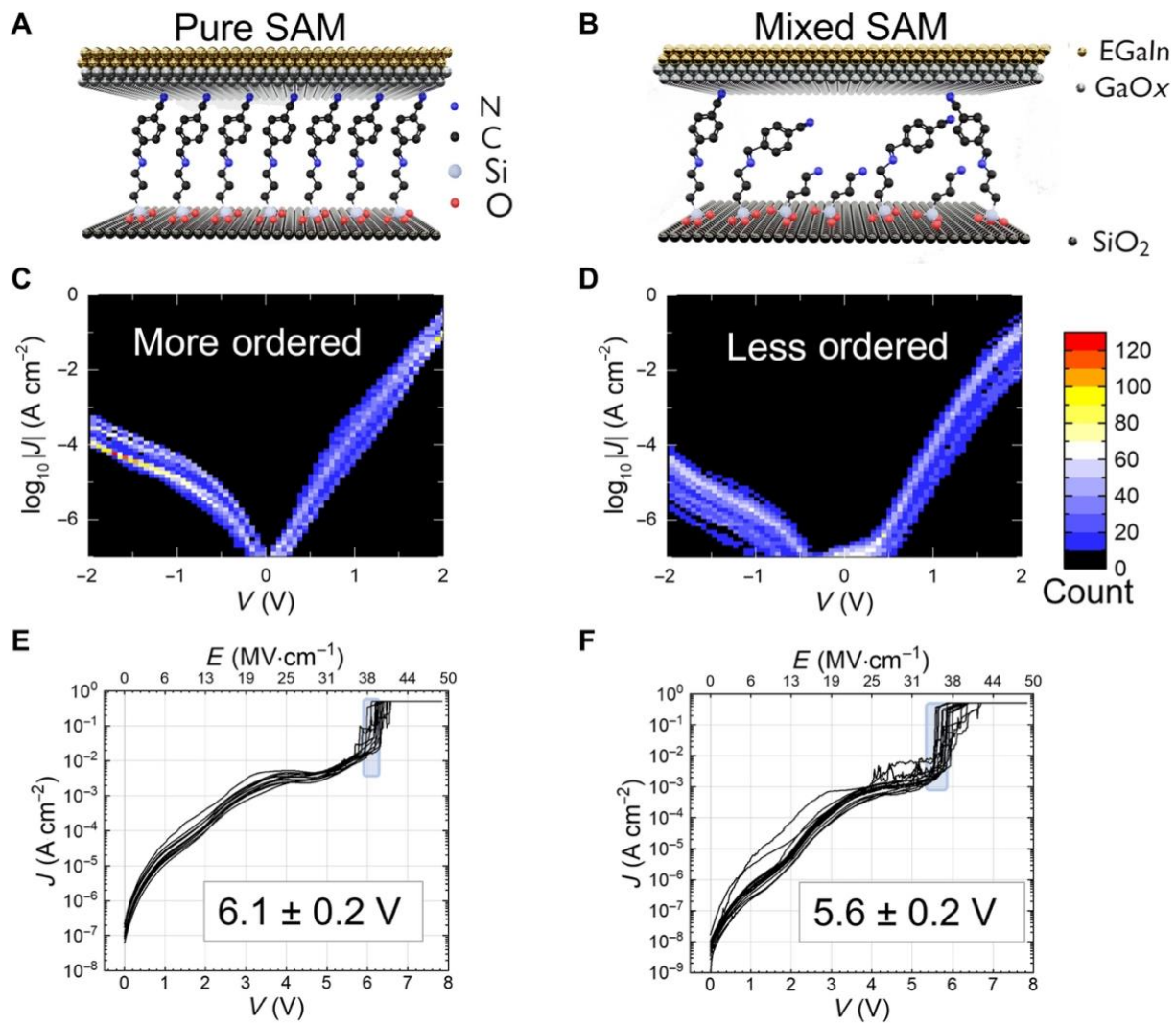

**Fig. 2. Uniformity in SAM films.** (**A**) Side view of the structure of molecular junctions consisting of CMPTM. (**B**) Side view of the structure of molecular junctions consisting of CMPTM co-assembled with APTES. (**C**) Multiple current density versus applied voltage curves in pure SAM sorted into 7000 two-dimensional (2D) bins. (**D**) Multiple current density versus applied voltage curves in mixed SAM films, sorted into 7000 2D bins. (**E**) Breakdown voltage measurements in pure SAMs. The average value is included in the inset. (**F**) Breakdown voltage measurements in mixed SAMs. The average value is included in the inset.

We found that the *J-V* curves for the mixed SAMs are spread over a wider range of values, suggesting that APTES introduces a greater degree of disorder upon mixing it in the rectifying SAMs. The lower conductance of the mixed SAMs, as it appears from Fig. 2D, also signals a change in molecular packing, gauche defects, packing density, film average thickness, and effective contact area (*21*, *22*). We eliminated the possibility of multilayer formation via physisorption or polymerization, which would lead to a more drastic decrease in conductance, as described later (*23*, *24*). Information on the density of pin holes within the SAM can be deduced by increasing an externally applied bias until the SAM breaks down, which is indicated by a spike in current, i.e., a sharp increase of at least an order of magnitude. Breakdown voltage measurements were taken on multiple spots for pure and mixed SAMs at a 70:30 (CMPTM:APTES) molar ratio (Fig. 2, E and F). The corresponding values were 6.1 ± 0.2 V and 5.6 ± 0.2 V, for the pure and mixed SAM, respectively. The higher breakdown voltage of the pure SAM film indicates that it has a lower density of pinholes than the mixed SAMs and, hence, is densely packed, with a higher degree of order, in agreement with the conclusion extracted from the spread in the *J-V* curves (*5*). The reduction in order of the SAM upon diluting it with another SAM has been documented in literature for other molecular rectifiers (*25–27*), but in all previous cases, it yielded a less efficient current rectification and, in some cases, the complete elimination of rectification even for low additive concentrations. This implies that an additional phenomenon occurs in our devices allowing for an increase in the strength of the current rectification, despite the reduction in order. We also ruled out the possibility of a chemical reaction between CMPTM and APTES being responsible for the changes found in the co-assembled SAMs (details are provided in Materials and Methods). In addition, APTES molecules were backfilled after CMPTM self-assembled to evaluate whether polymerization or hidden physisorbed additive molecules were increasing the rectifiers' performance through a change in leakage current due to the presence of multilayers. However, we found that the resulting samples were less than 50% efficient compared to pure SAMs of CMPTM, and eight times worse than the co-assembled CMPTM:APTES.

The surface topography of the single-component and mixed SAMs was investigated through static contact angle measurements. According to Cassie's law

$$\cos \theta_C = \sigma_1 \cos \theta_1 + \sigma_2 \cos \theta_2. \quad (1)$$

the wettability of a composite surface, quantified in the cosine of the static water contact angle, $\theta_C$, is the linear superposition of the cosine of the contact angle of its components ($\theta_1$, $\theta_2$) weighted by their surface coverages ($\sigma_1$, $\sigma_2$) (*28*). Strictly speaking, this theory is valid only if the components are of similar length. For the case when the molecules have different lengths, the shorter molecule is shielded from the top surface at low concentrations, which results in negligible changes in the contact angle (*27*). These deviations from Cassie's law, however, provide useful information on phase segregation and conformational defects present on the surface of the mixed SAM films and hence the evolution of long-range order. Once the concentration of the short molecule is sufficiently high (50 to 70%), its presence is detectable at the surface of the monolayer and the system obeys Cassie's law again. The evolution of the static water contact angle measured under ambient conditions on films of single component and systematically diluted SAM is shown in the graph presented in Fig. 3 in red; the calculated trend following Cassie's law is also included, in black, for reference. It can be observed that the value remains constant for small content of APTES due to the fact that the short molecules effectively integrate with the longer CMPTM molecules and thus will be hidden from the surface (regime *a* in Fig. 3). When the mixed SAMs are composed of 60% or more APTES, the contact angle value abruptly shifts to match the predicted trend of Cassie's law (regime *b* in Fig. 3). This result confirms that in the region that yielded the best rectification (30-40% APTES), the host and additive molecules form highly integrated SAMs, but the large difference in their lengths yield defects due to conformational changes that must occur in the long molecules to account for the unoccupied volume and possibly phase-segregated regions: Both effects lead to a reduction in order in the monolayers (*27*). The phase segregated domains, if present, are much smaller than the dimensions tested in the contact angle measurement, which are similar to the device size. The unimodal distribution histogram in Fig. 1E supports this conclusion since the rectifiers behave electrically as single-component, homogeneous SAMs. As expected, similar disruption in order has been observed upon mixing with triethoxy(propyl)silane, a nonrectifying molecule of similar size with APTES (fig. S6). However, unlike the case of CMPTM:APTES mixed SAMs, here the rectification of the mixed SAMs decreased by a factor of about 150% upon the reduction in order. These findings suggest that two competing processes occur upon mixing SAMs in our samples: (i) disruption of order within the monolayer and (ii) electronic coupling between the two different constituent molecules. In the case of triethoxy(propyl)silane, the electronic interactions are weak, and the reduced order prevails, leading to deterioration in device properties. In contrast, the interaction between the ($-COOCH_3$) group on the host molecule and the amino ($-NH_2$) group on APTES must be responsible for the observed enhanced current rectification, despite the fact that mixing introduces more disorder. We also tested mixed-SAM rectifiers in which CMPTM was co-assembled with *N*-(6-aminohexyl)aminomethyltriethoxysilane, a ($-NH_2$) substituted molecule of similar length in an attempt to enhance the rectification even further by maintaining the order in the SAM. The phase separation of the two SAMs in the absorbed monolayer, however, confirmed by the bimodal Gaussian distribution (two peaks) in the rectification histogram, inhibited the CT and thus the rectification remained unchanged (fig. S7).

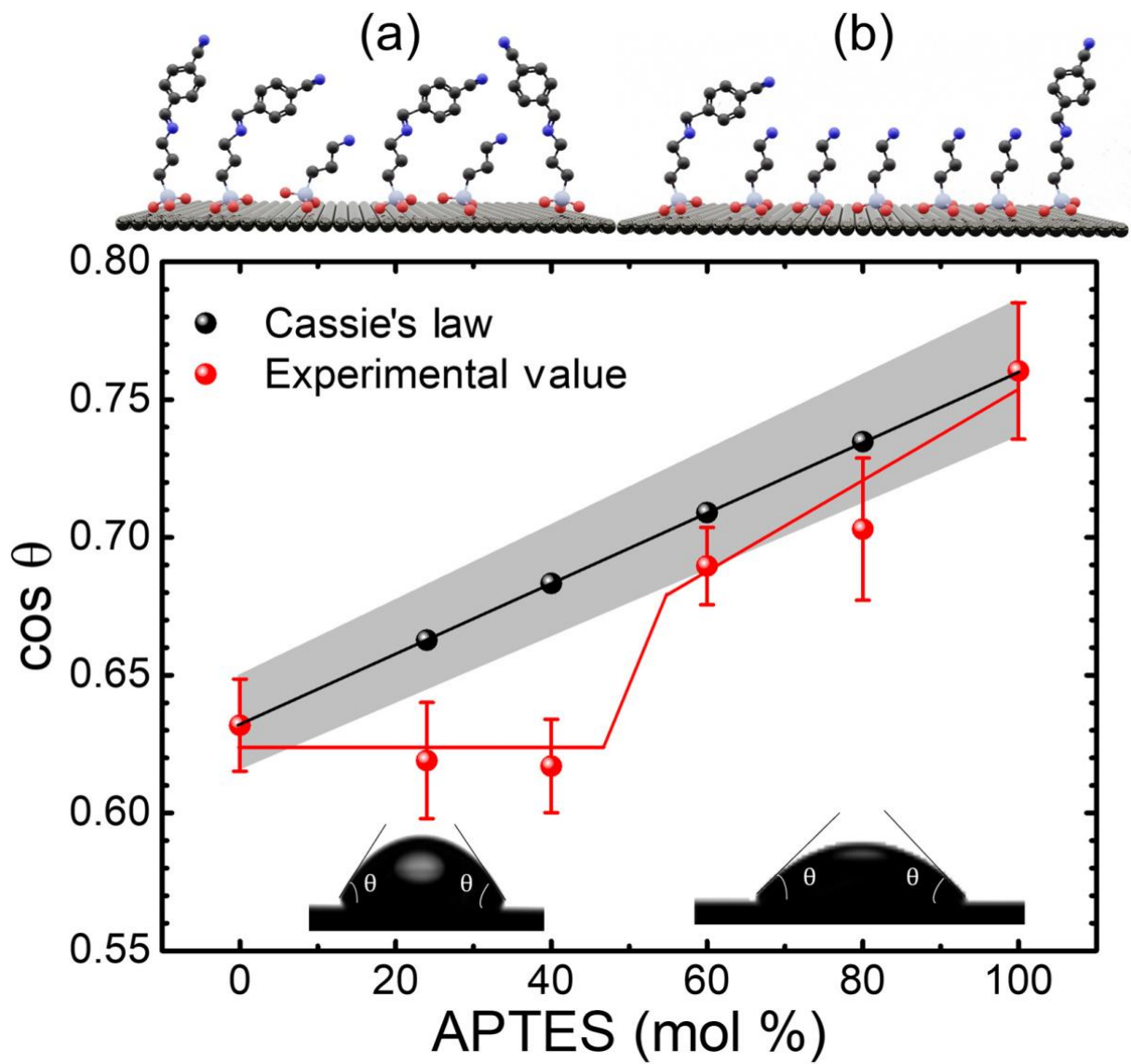

**Fig. 3. Cassie's law in mixed SAMs.** Contact angle measurements for mixed SAMs of CMPTM and APTES at various mixing ratios, confirming the presence of two distinct regions (red). The calculated points following Cassie's law (black).

Mixing SAMs has been recognized as a simple and effective method for fine-tuning the work function of a surface through altering the net dipole moment normal to the surface (*29–33*). The shift in work function, $\Delta\phi$, is a consequence of the fact that the intrinsic dipole of the SAM molecules changes the net surface potential, with the molecular density ($N$), orientation with respect to the substrate, and the strength of the dipole affecting the value and direction of the shift (*34*). $\Delta\phi$ is given by

$$\Delta\phi = eN\Delta\mu_\perp/\varepsilon. \qquad (2)$$

where $e$ is elementary charge, $\Delta\mu_\perp$ is the average dipole of the absorbed molecules perpendicular to the surface, and $\varepsilon$ is the dielectric constant (*30*), leading to a linear relationship between the concentration of an added diluent and the net change in work function. We evaluated the shift in the surface work function of the $SiO_2$ substrate upon SAM treatment using

Kelvin-probe measurements; results are shown in Fig. 4. Surface modification with CMPTM resulted in an increase in the work function by about 0.1 eV due to the fact that its electron-withdrawing terminal group (─COOCH$_3$) has the internal dipole pointing toward the electrode surface. In contrast, the electron-donating terminal group in APTES (─NH$_2$) has a dipole moment pointing in the opposite direction, which lowers the surface work function. The shift is very small in this case (−0.01 eV), most likely due to the fact that APTES molecules are tilted with respect to the surface normal (*35*). In the absence of any CT between the two SAMs and considering only the electrostatic interactions, a linear decrease in the work function shift following Eq. 2 is expected by increasing the APTES concentration, as predicted theoretically and shown experimentally in other systems (*36*, *37*); we depict this change schematically in violet in Fig. 4. The nonlinear relationship recorded between work function and APTES concentration is indicative of a change in the SAMs' net tilt angle to increase the component of the dipole moment normal to the surface and/or of electronic interactions such as partial CT between the APTES and the CMPTM molecules (*29*, *38*). Since we have unequivocally shown that the order in the mixed SAMs is reduced, increasing the net tilt angle of the SAM is highly unlikely. Therefore, we hypothesize that a partial CT ($\rho$) between the two molecules occurs. Such a coupling is facilitated by the fact that one SAM has an electron-withdrawing terminal group (─COOCH$_3$), while the other has an electron-donating terminal group (─NH$_2$). The CT enhances the electron density at the ─COOCH$_3$ end of the CMPTM molecule and hence increases its internal dipole moment, leading to a larger shift in the surface work function, in agreement with the experimental results included in Fig. 4. The highest degree of CT, inferred from the largest work function shift, coincides with the APTES concentration that yielded the highest current rectification (Fig. 1C).

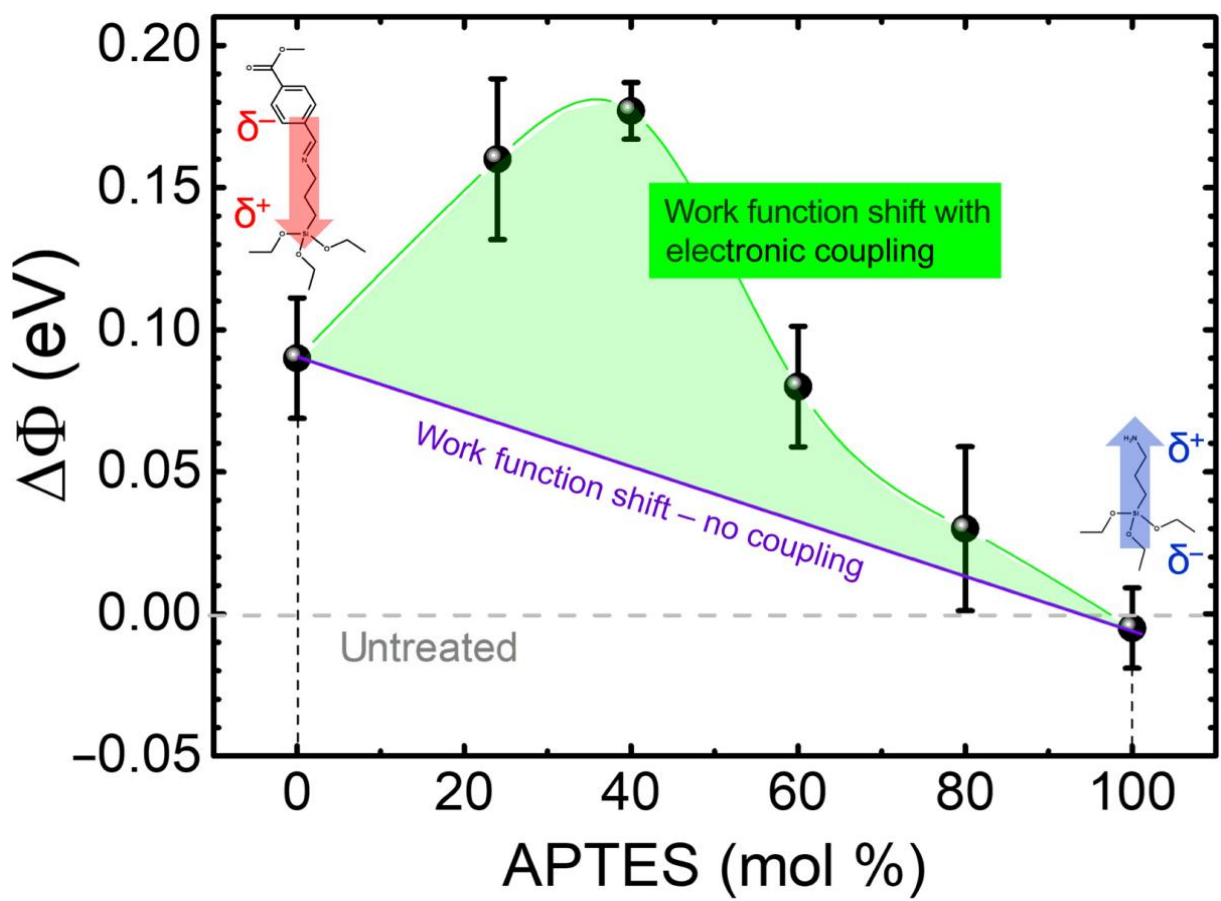

**Fig. 4. Work function shift in substrates treated with mixed SAMs.** Dependence of the work function shift of the SiO$_2$ substrate on the CMPTM/APTES ratio. The purple line represents the estimated shift in the absence of CT between the two SAMs. The internal molecular dipoles are indicated in red and blue, respectively.

This CT state characterized by partial ionicity ρ is therefore responsible for the new properties observed in our molecular rectifiers. The degree of CT depends on the strength of coupling between the electron-acceptor and electron-donor functional groups, which, in turn, is controlled by the chemical nature of the two substituents and their spatial overlap, similar to the case of doping organic semiconductor films or CT salts (*15*). Since the latter is directly related with the degree of order in the SAM, future studies focusing on the interplay between processing and morphology within the mixed SAMs are imperative to gain a better control of these processes.

Having established that CT can occur in our mixed SAMs, we now turn to a discussion on how it affects the mechanism responsible for the current rectification in the molecular diodes. The similarity in the work functions of the silicon and EGaIn electrodes (−4.65 and −4.2 eV, respectively) excludes the possibility of a Schottky mechanism (*19*, *39*). In molecular diodes based on the pure SAM, the delocalization of the lone pair of electrons from the ─COOCH$_3$ tail of the molecules allows for a coupling between them and the top EGaIn electrode, which leads to Fermi level pinning. This phenomenon is consequential to the position of the frontier molecular orbitals and gives rise to efficient charge transport for one bias polarity and inefficient charge transport for the opposite polarity, as it has been shown in past studies (*11*).

In the mixed SAMs, doping of the molecular rectifier molecule adds an additional level of complexity. Adopting the vocabulary specific to conventional p-n junction diodes, Fig. 5 depicts the energy-level diagrams under forward (+2 V) and reverse (−2 V) bias applied at the EGaIn electrode, with the silicon electrode being grounded. When a positive voltage is applied, the molecule/electrode coupling raises the energy levels in the SAM, placing the lowest unoccupied molecular orbital (LUMO) in the transmission window, similar to the case of pure SAM (Fig. 5A). In addition, the energetic landscape allows for efficient electron transfer from the donor additive SAM to the rectifying molecule, similar to n-type doping. Transport occurs via a combination of nonresonant tunneling (dashed yellow line) and the resonant tunneling (solid yellow line), with the second being highly efficient, and thus responsible for the high currents. The partial CT increases the conductivity (see the inset in Fig. 1B), despite the fact that the morphology becomes less optimal. When a negative bias is applied, the LUMO shifts outside the transmission window and transport takes place via nonresonant tunneling (Fig. 5B). The shift in the energy levels of the rectifier molecules also alters the energy-level alignment with the additive molecules, which do not make intimate contact with the top EGaIn electrode, and thus do not undergo electrical coupling. Hence, on the basis of the energy-level diagram, CT cannot occur now. Collectively, these phenomena yield lower conductivity in the reverse bias regime than the case of forward bias and thus an efficient current rectification. An important point is that the underpinning performance enhancement upon CT is not limited to the case where the rectification mechanism involves Fermi level pinning caused by the strong molecular/electrode coupling: Even for a system with low or negligible interaction between the rectifier SAM and the contacts, CT leads to an increase in the internal dipole moment of the molecule, making the current rectification more efficient (*19*).

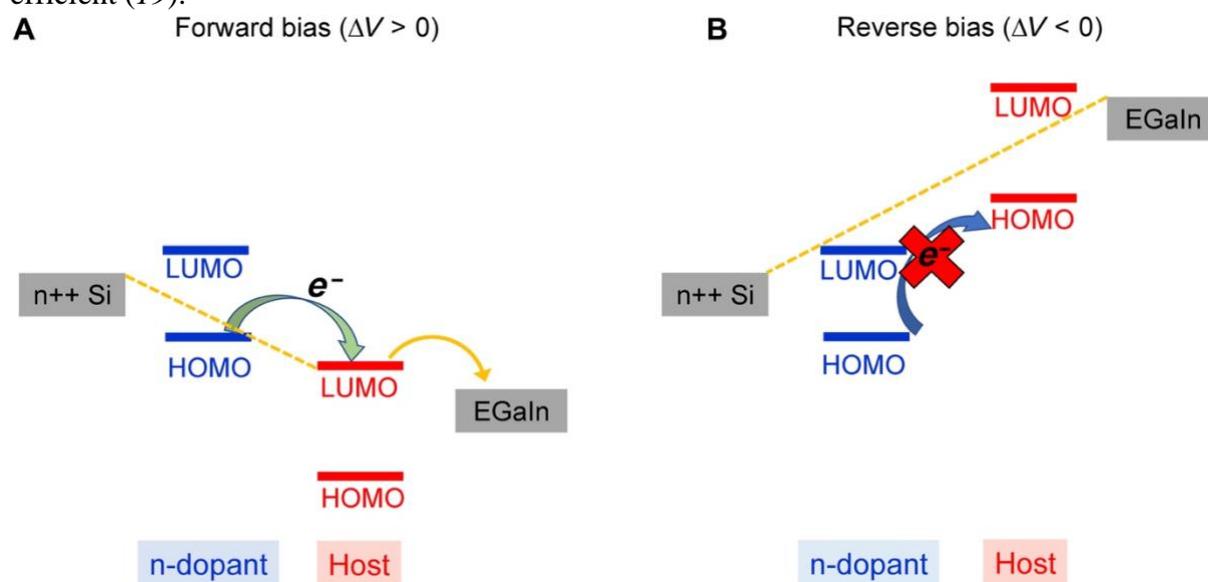

**Fig. 5. Energy diagram in a film consisting of mixed SAMs of CMPTM and APTES.** (**A**) Energy levels when a positive bias is applied. (**B**) Energy levels when a negative bias is applied.

We further investigated the D/A CT as a strategy for enhancing the performance of molecular diodes in a series of SAMs that were designed and synthesized to contain functional groups with various electron affinities in order to tune the strength of interaction with the additive.

Devices consisting of both single-component and mixed SAMs co-assembled with the same additive, i.e., APTES, at a 70:30 ratio, have been evaluated; results are shown in Fig. 6A. Two groups of rectifying molecules were tested: The group presented on the left, in blue, includes (*E*)-1-(4-carbomethoxy-3-methyl-phenyl)-*N*-(3-(triethoxysilyl)propyl)methanimine (CMMPTM), (*E*)-1-(4-cyanophenyl)-*N*-(3-(triethoxysilyl) propyl)methanimine (CPTM), and CMPTM, all molecules that contain an electron-accepting functional group, as quantified by a positive value of the Hammett constants (table S1) (*40*). In the framework of our hypothesis, CT from the donor ─NH$_2$ in APTES additive to the electron-accepting termini of the rectifying SAMs can occur in this case and, consequently, an enhancement in rectification strength is expected. An increase in rectification was observed upon mixing these SAMs with APTES, as shown in blue bars in Fig. 6A (the value is normalized with respect to the rectification of the pure SAM). In contrast, for the second group, consisting of SAMs with electron-donor functional groups such as (*E*)-1-(4-methoxyphenyl)-*N*-(3-(triethoxysilyl)propyl)methanimine (MPTM), (*E*)-1-(4-(methylthio)phenyl)-*N*-(3-(triethoxysilyl)propyl)methanimine (MTPTM), and (*E*)-1-phenyl-*N*-(3-(triethoxysilyl)propyl)methanimine (PTM) (described by negative Hammett constants, complete names provided in Materials and Methods) and shown in red on the right, in black bars, the rectification ratios either decreased or suffered negligible changes due to the fact that CT did not occur. In summary, as expected, the rectification of the mixed SAMs increased relative to the pure SAMs only when a combination of molecules with strongly polar acceptor and donor end-groups was co-assembled. We do not exclude, however, the idea that the coupling between the SAMs and the electrode can also play a role, but we expect that to affect only the rectification strength of the host SAM, not the relative change. These observations confirm that the observed effects are more general, and the mixed SAMs provide an opportunity for controlling the electrical performance of molecular rectifiers by providing access to states inaccessible by pure SAM tunneling junctions.

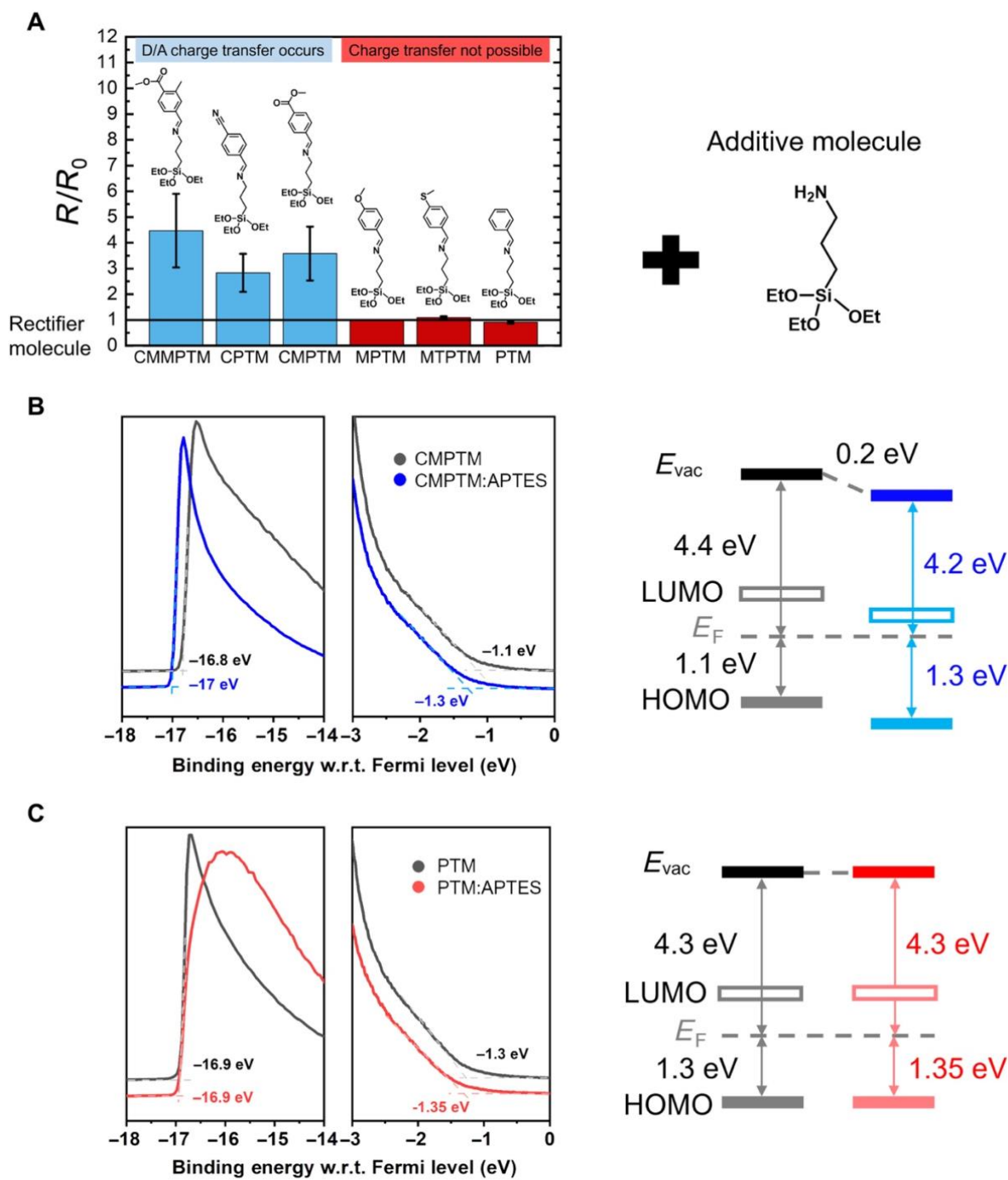

**Fig. 6. Spectroscopic evidence for doping in SAMs co-assembled with APTES.** (**A**) The normalized difference of $R$ between pure and mixed SAMs when mixed with APTES (70:30) for molecules with accepting terminal groups (blue) and donating terminal groups (red) determined by Hammett constants. Chemical structures of all rectifying compounds: CMMPTM, CPTM, CMPTM, MTPTM, MPTM, PTM, and APTES. (**B**) UPS measurement on the CMPTM SAM (black) and the CMPTM:APTES 70:30 mixed SAM (blue) along with the corresponding energy-level alignment confirming a deeper HOMO and a smaller work function as a result of CT. (**C**) UPS measurement on the PTM SAM (black) and the PTM:APTES 70:30 mixed SAM (red) showing no relevant changes in the energy-level alignment. $E_{vac}$ represents the vacuum energy, and $E_F$ denotes the Fermi energy.

To obtain direct evidence for the presence/absence of CT for molecules with accepting/donating terminal groups, UPS measurements were performed on one representative set of pure and mixed SAMs from each category to evaluate the changes occurring in the band alignment. In the case of mixed CMPTM/APTES, we observed that the secondary photoemission onset, which provides information about the work function of the samples, is shifted by 0.2 eV as compared to pure CMPTM (Fig. 6B). A similar shift is observed in the position of the highest occupied molecular orbital (HOMO) with respect to the Fermi level, thus confirming a reduction in the work function due to a CT from the APTES additive to the host SAM molecule, in agreement with the Kelvin probe results depicted in Fig. 4. The larger HOMO-to-$E_F$ (Fermi level) offset is consistent with n-type doping. In contrast, both the energetic position of the photoemission onset and that of the HOMO level of the pure PTM and that of the PTM/APTES-mixed SAMs vary only by 0.05 eV, well within the experimental error of UPS (Fig. 6C), thus confirming that no CT takes place in this case, in agreement with our interpretation of enhancement of the molecular rectifier performance upon CT.

The results presented so far collectively suggest that mixing SAMs with donor termini with SAMs with acceptor termini provides a simple and reliable route to enhance the performance of molecular diodes. Next, we incorporated the rectifiers in basic electronic circuits in which we replaced the silicon diode with our high-performance molecular diodes. In addition to the low-power consumption, the small size of molecular diodes makes them ideal candidates for rectifier circuits due to their inherently low resistance. Since the distance between the two contacts is controlled by the length and orientation of the SAM molecule, the definition of a short channel necessary to increase the cutoff frequency is straightforward in these devices, without the need of complicated lithography processes. To test the compatibility of our molecular diodes with real circuits, we incorporated them into simple proof-of-concept AC/DC rectifiers that convert an alternating current (AC) input voltage $V_{in}$ to a direct current (DC) output voltage $V_{out}$ (Fig. 7). The schematic cross section of the molecular diode is shown in Fig. 7A, along with typical current-voltage characteristics in Fig. 7B, while in Fig. 7C, we present the circuit diagram of the DC rectifier. The smoothing capacitor had a capacitance of $C = 100$ nF, while the value of the load resistor was $R = 1$ megohm. Figure 7D displays the AC input voltage of a frequency of 100 kHz and the resulting DC output voltage; it can be clearly observed that the ripple voltage is negligible as a result of our high-quality molecular diode.

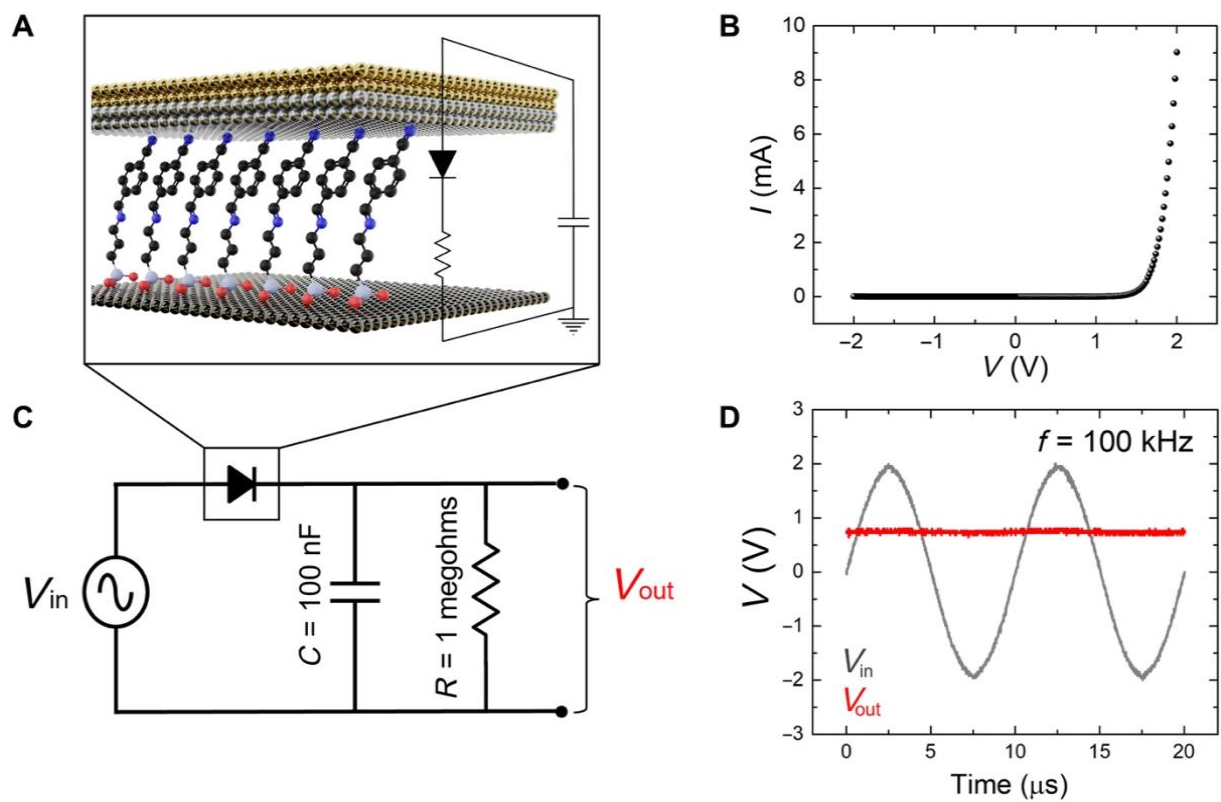

**Fig. 7. Basic circuits with molecular rectifiers consisting of mixed SAMs.** (**A**) Cross-sectional schematic of the molecular rectifiers. (**B**) Typical current-voltage characteristics of the molecular rectifiers. (**C**) AC/DC rectifier circuit diagram. (**D**) Input and output voltages corresponding to the circuit presented in (C).

**Discussion**

We introduced a new concept for tuning the properties of molecular rectifiers, which relies on CT processes in co-assembled SAMs, where one SAM contains an electron-accepting terminal group and the other contains an electron-donor terminal group. This process grants a high level of control over the electronic properties of the monolayers and allows for an unprecedented efficiency in current rectification. We found that the device performance correlates positively with the efficiency of CT, leading to rectification ratios approaching $10^4$, on par with those obtained in SAMs deposited on metallic electrodes, but with the additional advantages of allowing a straightforward integration with current silicon technologies and cost efficiency due to reduced complexity in processing. Our results not only have implications into the technology of molecular electronic devices, but they represent the first demonstration of CT processes in SAMs. Further studies on mixtures of SAMs with predetermined offsets in their frontier orbitals so that the degree of CT is systematically modified in these simple systems could provide the much-needed understanding on donor-acceptor interactions, thus guiding the process of doping in more complex electronic materials systems used in photovoltaics, light-emitting diodes, and thermoelectrics. These systems have the potential to move the field forward in a substantial way, by opening up the prospect of the discovery of unexpected material properties arising from the intermolecular interactions between the D and A termini of the molecules within the mixed SAMs. These novel functionalities, governed by the degree of CT, are not manifested in the parent compounds, thus offering a platform for novel material discovery, with an extraordinary flexibility of combining donors and acceptors. Given the soft nature of the intermolecular interactions, the main challenge lies in maintaining the degree of order in the rectifying SAM upon co-assembly with a dopant SAM. Therefore, to access the full potential of this strategy, development of methods that could preserve the mixed SAM degree of order is imperative. The anticipated long-term impact is that molecular devices may one day be integrated with mature, conventional semiconductor technologies to enable them with complementary functionalities like chemical and biological sensing, by only adding a very small fraction of the total cost.

## Materials and Methods

### Device fabrication
Highly n-doped silicon wafers (resistivity of 0.001 to 0.005 ohm cm) with a native oxide layer were cut in 1 × 1 cm substrates, then cleaned for 10 min in an acetone bath at 85°C, and immediately followed by an acetone and isopropyl alcohol (IPA) rinse. Then, the substrates were immersed in an IPA bath at 85°C for 10 min, rinsed with IPA, and dried with a stream of nitrogen. Last, the wafers were exposed to ultraviolet ozone for 10 min, thoroughly rinsed with deionized water, and dried. The substrates were then brought into a glovebox [<0.1 parts per million (ppm) $H_2O$, <0.1 ppm $O_2$], where each of them was placed in a jar containing a 7 mM solution of the SAM (or stoichiometric SAM mixtures) in room-temperature chloroform for 16 to 24 hours. For mixed SAMs, the molar ratio of additive to host molecules was adjusted between 0 and 100%. The D/A SAM ratio was controlled by tuning the stoichiometry of the mixed SAM solution, but small deviations from these values can occur in the absorbed monolayer due to the various processes governing the co-adsorption process (*41*, *42*). After self-assembly, the SAMs were thoroughly rinsed with chloroform and IPA and then dried. The rectifying SAMs (illustrated in Fig. 6) consisted of CPTM, CMPTM, PTM, CMMPTM, MPTM, and MTPTM.

### Electrical characterization
The SAMs were electrically characterized in an ambient probe station in a metal/SAM/metal sandwich configuration. The EGaIn top contact was prepared by dipping a 0.5-mm-diameter tungsten tip into EGaIn, forming a malleable cone. The silicon substrates played the role of the bottom contacts and were grounded, while an external potential was applied at the top EGaIn electrode. Several molecular junctions were formed in this way on one 1 × 1 cm substrate, and each molecular junction was measured using an Agilent 4155C semiconductor parameter analyzer. At least 200 measurements, resulting from a minimum of five samples, were made for each type of sample. The rectification ratio, $R$, was calculated as the ratio between the current density $J$ at forward bias $V_{fwd}$ and the current density at reverse bias $V_{rev}$. The current density was obtained by dividing the measured current by the geometrical area of the EGaIn probe tip. The voltage bias used in this report is ±2 V. Breakdown voltages of the SAMs were averaged for 20 different molecular junctions by sweeping an external potential from 0 to −8 V. Bias stress stability tests were performed by sweeping the voltage from −2 to 2 V for 100 cycles on the same junction under ambient conditions with no wait time between measurements.

**Work function determination**
Single-component and mixed SAMs of CMPTM and APTES assembled onto a Si substrate terminated with a very thin native oxide layer were analyzed using a Trek model 325 electrostatic voltmeter configured for Kelvin probe measurements. The Kelvin probe was calibrated using highly ordered pyrolytic graphite (HOPG). The work function, $\phi$, of the untreated substrate and of the substrate upon each treatment was determined by the expression

$$\phi = -e(V_{SAM} - V_{HOPG}) + \phi_{HOPG}. \qquad (3)$$

where $e$ is elementary charge, $V_{SAM}$ is the signal measured on the Si or SAM/Si substrate, $V_{HOPG}$ is the signal from HOPG, and $\phi_{HOPG}$ is the work function of HOPG ($\phi_{HOPG}$= 4.48 eV) (*43*). At least three measurements were taken for each treatment, and the results were compared to the measured work function measured on the clean, untreated substrates.

**Contact angle measurements**
Static contact angle measurements were taken in a Ram'e-Hart Model 200 Contact Angle Goniometer using 3-μl deionized water droplets placed near the center of 1 × 1 cm SAM film. At least five contact angle measurements were used for each pure and mixed SAM combination.

**CMPTM stability**
A small shard of silicon substrate was deposited into a 2-ml nuclear magnetic resonance (NMR) sample tube along with a 7 mM CDCl$_3$ solution of CMPTM with 30% APTES, and HNMR spectra of the mixture were taken after 4 and 24 hours to screen for the presence of side reactions or reversion to starting material (fig. S8). The HNMR spectrum of APTES is also included in this figure for comparison. We saw no measurable increase in the peak at 10 ppm corresponding to the starting aldehyde used to synthesize CMPTM, and we saw no notable side product formation as indicated by the consistency of the substrate containing spectra with the spectrum of pure CMPTM. We can conclude from these data that no undesired side reaction or decomposition of CMPTM is occurring in the deposition and the final device consists only of CMPTM and APTES.

**Aromatic-N-(3-(triethoxysilyl)propyl)methanimine synthesis**
All aromatic-*N*-(3-(triethoxysilyl)propyl)methanimines used in this work were obtained from the reactions of the appropriate benzaldehydes with APTES as described previously in more detail in the Supplementary Materials (*11*, *44*). All methanimines used in this work were characterized by $^1$H and $^{13}$C NMR spectroscopy and high-resolution mass spectrometry, and the NMR spectra of these compounds are included in the Supplementary Materials (figs. S9 to S23).

**Ultraviolet photoemission spectroscopy**

The samples were transferred to an ultrahigh vacuum chamber (ESCALAB 250Xi), with a base pressure of $2 \times 10^{-10}$ mbar, for UPS measurements. UPS measurements were performed using a double-differentially pumped He gas discharge lamp emitting He I radiation (hν = 21.22 eV) with a pass energy of 2 eV and a bias of −5 V to ensure secondary electron onset detection. The UPS spectra are shown as a function of the binding energy with respect to the Fermi energy. The energy edge of the valence band features is used to determine the valence band level position with respect to the Fermi level.

**Acknowledgments**
We thank J. E. Anthony, K. Graham, and S. Banks for useful discussions.
**Funding:** The work at Wake Forest University was partially supported by the NSF under grant ECCS-1810273. The work at TU Dresden received funding from the European Research Council (ERC) under the European Union's Horizon 2020 research and innovation programme (ERC grant agreement no. 714067, ENERGYMAPS). O.D.J. and Y.V. thank the NSF and the Deutsche Forschungsgemeinschaft for funding in the framework of a joint NSF-DFG grant in Advanced Manufacturing (NSF: CMMI-2135937 and DFG: VA 991/8-1).
**Author contributions:** R.P.S., M.E.W., and O.D.J. conceived the project and designed the experiments. R.P.S., E.C.-T., and R.W.B. fabricated the samples, executed the electrical and morphological characterization, and analyzed the data under the supervision of O.D.J. J.T.M. synthesized the SAMs and performed the chemical characterization of the new compounds under the supervision of M.E.W. Y.J.H. and Y.V. performed the UPS measurements. R.P.S. and O.D.J. wrote the manuscript. All authors contributed to the discussions and edited the manuscript.